%% file: main.tex
\documentclass[review]{elsarticle}
\usepackage[utf8]{inputenc}
\usepackage[T1]{fontenc}
\usepackage{geometry}
\usepackage{xurl}
\usepackage{enumitem}
\usepackage{placeins}
\usepackage{url}
\usepackage{booktabs}
\usepackage{longtable}
\usepackage{array}
\usepackage{tabularx}
\usepackage{ragged2e}
\journal{International Journal of Critical Infrastructure Protection}
\begin{document}

\begin{frontmatter}

\title{From Backup Restoration to Minimum Viable Factory Recovery: A Systematization of Ransomware Recovery in Manufacturing Systems}

\author[inst1]{Chun Yin Chiu\corref{cor1}}
\ead{chun\_yin.chiu@kcl.ac.uk}
\cortext[cor1]{Corresponding author}
\affiliation[inst1]{organization={King's College London, Strand, London WC2R 2LS, United Kingdom}}

\begin{abstract}
Ransomware recovery in critical manufacturing infrastructure is not only a backup-restoration problem. Production capability depends on coupled information-technology, operational-technology, physical-process, quality, logistics, identity, and supplier systems. After ransomware, a plant may rebuild servers yet remain unable to schedule work, authenticate operators, trust engineering workstations, release product, reconnect OT assets, or coordinate suppliers. This paper reframes manufacturing ransomware recovery as a critical-infrastructure continuity and interdependency problem. We conduct a PRISMA-guided multivocal review of academic literature, standards and government guidance, threat frameworks, public incident material, and verified full-text/source-page evidence anchors. The review identifies nine evidence-backed recovery failure modes: dependency blindness, untrusted restore point and backup over-trust, identity trust collapse, lack of proof-of-recovery, unsafe OT reconnection, segmentation assumption failure, capability mismatch, unmanaged degraded operation, and supplier dependency failure. We then introduce Minimum Viable Factory Recovery (MVF Recovery): the smallest safe, trusted, and operationally meaningful production capability that can be resumed under current dependency, evidence, identity, data, network, OT, and supplier constraints. MVF Recovery is an analytical objective rather than a claim of full recovery, implementation, or safety certification. The paper derives a recovery lifecycle and benchmarking directions as secondary outputs. The contribution is an evidence-calibrated foundation for capability-centric ransomware recovery in critical manufacturing infrastructure.
\end{abstract}


\end{frontmatter}

\input{sections/01_introduction.tex}
\input{sections/02_background_and_related_work.tex}
\input{sections/03_methodology.tex}
\input{sections/04_evidence_backed_recovery_failure_taxonomy.tex}
\input{sections/05_minimum_viable_factory_recovery.tex}
\input{sections/06_evidence_based_recovery_lifecycle.tex}
\input{sections/07_benchmarking_directions.tex}
\input{sections/08_discussion_and_limitations.tex}
\input{sections/09_conclusion.tex}
\section*{CRediT authorship contribution statement}
Chun Yin Chiu: Conceptualization, Methodology, Investigation, Data curation, Formal analysis, Writing - original draft, Writing - review \& editing, Visualization.

\section*{Declaration of competing interest}
The author declares that there are no known competing financial interests or personal relationships that could have appeared to influence the work reported in this paper.

\section*{Data availability}
The review artefacts supporting this manuscript are included as supplementary material. They include a database-level search strategy and export-count log, PRISMA-style flow counts, aggregate screening decisions, representative full-text/source-page verification anchors, an adjudicated failure-mode evidence summary, a methodological consistency check, a cited-source evidence matrix, and candidate research gaps. Access-restricted publisher PDFs and third-party copyrighted reports are not redistributed.

\section*{Funding}
This research did not receive any specific grant from funding agencies in the public, commercial, or not-for-profit sectors.
\FloatBarrier
\input{sections/10_references.tex}

\end{document}

%% file: sections/01_introduction.tex
\section{Introduction}

Critical manufacturing infrastructure differs from generic enterprise computing because digital recovery is inseparable from physical production. A factory does not become recovered merely because encrypted servers are rebuilt or backups are restored. Production capability depends on a tightly coupled chain of identity services, engineering workstations, manufacturing execution systems (MES), historians, quality databases, enterprise resource planning (ERP), logistics platforms, supplier interfaces, operators, maintenance procedures, safety constraints, and OT assets such as HMIs, PLCs, sensors, actuators, and SCADA components. Ransomware can therefore interrupt production without directly damaging every machine on the factory floor: it can remove visibility, block scheduling, corrupt trust in engineering systems, disable quality release, prevent safe OT reconnection, or break supplier/customer coordination.

This coupling makes manufacturing recovery a specific critical-infrastructure problem. First, manufacturing is time-sensitive and sequence-sensitive: upstream material, production scheduling, quality inspection, packaging, shipping, and customer commitments are interdependent. Second, manufacturing involves cyber-physical safety constraints that are not present in ordinary office IT. Systems must often be restarted in a known safe state, with controlled network paths and validated operator access. Third, factories commonly combine legacy OT, modern IT, vendor-maintained equipment, remote access, and brownfield IIoT gateways, creating recovery dependencies that cross organizational and technical boundaries. Fourth, partial recovery can be meaningful but risky: a plant may operate manually or in degraded mode, but such operation must still be safe, auditable, and capable of producing valid output. Finally, manufacturing disruptions cascade into customers, suppliers, critical sectors, and regional economies. These properties justify treating ransomware recovery in manufacturing as more than enterprise incident response applied to a different sector.

Public incidents in manufacturing and adjacent critical-infrastructure sectors illustrate the gap between asset restoration and capability restoration. Reported events have affected production scheduling, enterprise services, manual operations, logistics, supplier coordination, customer service, and staged restart. Some organizations resumed only constrained or manual operations; others experienced supply-chain effects after a supplier or business-system disruption. These cases show that the relevant recovery question is not simply whether encrypted machines can be rebuilt. It is whether a minimum safe, trusted, and operationally meaningful production capability can be re-established.

Existing research provides important foundations but does not fully answer this question. Ransomware literature has produced malware analyses, detection methods, backup mechanisms, and response guidance [1]--[6]. ICS/OT research has examined vulnerabilities, segmentation, safety constraints, and cyber-physical attack paths [7]--[15]. Critical-infrastructure research has emphasized interdependency, cascading failure, and continuity under degraded performance [16]. Digital forensics and forensic-readiness research has shown the importance of evidence preservation, clean-state reconstruction, and recovery support [17]--[19]. Supply-chain studies show that ransomware disruption can propagate across inter-firm relationships and that recovery depends on both internal capabilities and partner coordination [20]--[22]. ICS incident-learning and response/recovery framework papers in this journal have shown the value of extracting operational lessons from public incidents, standards, and practitioner-facing guidance [23], [24]. Yet these streams rarely converge on the post-ransomware recovery decisions faced by manufacturing operators: what should be restored first, which restore points are trustworthy, which identities and credentials can be reused, when OT assets may be reconnected, and what evidence is sufficient to resume production.

This paper addresses that gap by reframing ransomware recovery in critical manufacturing infrastructure as a capability-restoration problem. The central argument is that recovery fails when organizations optimize for restored assets while overlooking production dependencies, trust state, evidence quality, OT reintegration risk, degraded operations, and supplier coupling. We use the term \textbf{Minimum Viable Factory Recovery} (MVF Recovery) to describe the smallest safe set of trusted systems, identities, data, network paths, operational procedures, and external dependencies required to resume constrained but valid production after ransomware disruption. MVF Recovery is used as an analytical recovery objective, not as a claim that full recovery or formal safety certification has been achieved. MVF Recovery differs from generic business-continuity objectives because it treats production resumption as a constrained cyber-physical trust decision: the recoverable capability must satisfy dependency consistency, identity trust, clean-state evidence, OT reintegration safety, degraded-mode governance, and supplier feasibility simultaneously.

The paper is guided by four research questions:

\input{tables/research_questions_v2_4.tex}

We support these questions through a PRISMA-guided multivocal review of academic literature, government and standards guidance, threat frameworks, public incident material, and verified full-text/source-page evidence anchors. The review is evidence-calibrated: it does not claim complete full-text verification of every retained record or population-level incident prevalence. Instead, it identifies source-supported recovery failure modes and distinguishes them from under-reported candidate patterns. High-impact taxonomy claims are anchored to accessible full text, official source pages, or author-supplied full-text documents; access-restricted records are retained only as background or inferred support unless separately verified.

The paper makes two primary contributions. First, it presents an evidence-backed taxonomy of nine ransomware recovery failure modes specific to critical manufacturing infrastructure: dependency blindness; untrusted restore point and backup over-trust; identity trust collapse; lack of proof-of-recovery; unsafe OT reconnection; segmentation assumption failure; capability mismatch; unmanaged degraded operation; and supplier dependency failure. Second, it defines MVF Recovery as a capability-centric recovery objective linking dependency modelling, trust reconstruction, forensic evidence, safe OT reintegration, and constrained production resumption. We then derive two practical outputs: a compact evidence-based recovery lifecycle and benchmarking directions for future recovery evaluation. The benchmarking contribution is deliberately framed as a blueprint and research agenda rather than a released executable benchmark.

The rest of the paper is organized as follows. Section 2 positions the work against ransomware, ICS/OT security, critical-infrastructure resilience, digital forensics, and supply-chain recovery literature. Section 3 describes the review design, source-level eligibility assessment, full-text/source-page verification, coding validation and conservative claim calibration. Section 4 presents the taxonomy. Section 5 defines MVF Recovery. Section 6 develops the recovery lifecycle. Section 7 outlines benchmarking directions. Section 8 discusses implications, limitations, and research gaps. Section 9 concludes.

%% file: tables/research_questions_v2_4.tex
{\small
\begin{longtable}{@{}>{\RaggedRight\arraybackslash}p{0.45\textwidth} >{\RaggedRight\arraybackslash}p{0.49\textwidth}@{}}
\caption{Research questions and role in the synthesis.}\label{tab:research-questions}\\
\toprule
Research question & Purpose in this paper \\
\midrule
\endfirsthead
\toprule
Research question & Purpose in this paper \\
\midrule
\endhead
\textbf{RQ1.} What recovery failure modes are evidenced in ransomware-related manufacturing, ICS/OT, and critical-infrastructure literature? & Derive the evidence-backed taxonomy in Section 4. \\
\textbf{RQ2.} How do these failure modes show that asset restoration is insufficient for manufacturing recovery? & Establish the capability-centric recovery argument. \\
\textbf{RQ3.} What minimum capability objective can guide safe, trusted, and operationally meaningful post-ransomware recovery? & Define MVF Recovery in Section 5. \\
\textbf{RQ4.} What evaluation dimensions are needed for future benchmarking of manufacturing ransomware recovery? & Derive the lifecycle and benchmarking directions in Sections 6 and 7. \\
\bottomrule
\end{longtable}
}

%% file: sections/02_background_and_related_work.tex
\section{Background and Related Work}

\subsection{Ransomware recovery beyond malware removal}

Ransomware research has historically emphasized attack behaviour, detection, prevention, and data restoration. Malware-oriented studies classify ransomware families, encryption strategies, key management, propagation, and extortion models [1]--[3], [19]. Storage and recovery research has proposed mechanisms such as firmware-level or file-level recovery to reduce reliance on conventional backups [25], [26]. Government and national guidance emphasizes tested offline backups, incident response planning, clean restoration, credential resets, logging, and containment [4]--[6], [27]. These contributions are necessary but incomplete for manufacturing recovery. They often focus on whether data or systems can be restored, whereas a factory must also restore a safe and coordinated production capability.

Ransomware recovery is particularly difficult because it occurs under uncertainty. Responders may not know when the attacker first entered, which credentials were stolen, whether backups were reached before encryption, whether recovery infrastructure is trustworthy, or whether recovered systems will reinfect clean assets. Technical data-restoration methods are useful, but they do not decide whether a manufacturing execution system can be used to schedule production, whether an engineering workstation can reconnect to controllers, or whether a recovered quality database is valid enough to release product. This paper therefore treats ransomware recovery as a decision problem in which restoration actions must be evaluated against production missions and evidence constraints.

\subsection{ICS/OT constraints in manufacturing recovery}

Industrial and cyber-physical environments introduce recovery constraints that do not arise in ordinary enterprise IT. OT systems monitor and control physical processes; failures can affect safety, quality, equipment, environment, and service continuity [8]. ICS ransomware and IIoT studies show that attacks can target supervisory computers, PLCs, SCADA environments, edge gateways, or the IT/OT bridge, affecting visibility, control, and production operations [13], [14], [28]--[30]. Energy-delivery analysis further illustrates that ransomware may not need to cause physical destruction to be operationally serious: inhibiting real-time data access, operator commands, or situational awareness can still force shutdown or degraded operation [15].

These constraints make recovery sequencing central. A recovered HMI, historian, engineering workstation, or domain controller may be technically online but unsafe to reconnect if its state, credentials, network path, or process impact is untrusted. Conversely, some factories may need to resume limited production before full enterprise restoration, using manual workarounds or degraded modes. This requires recovery objectives that explicitly account for safety, trust, evidence, and dependency constraints.

\subsection{Critical infrastructure interdependency and continuity}

Critical-infrastructure literature emphasizes interdependent processes, cascading disruption, and minimum acceptable service. Cyber-resilience can be understood as the ability to absorb and adapt to cyber incidents while sustaining system process continuity, possibly at a degraded level [16]. This view is well aligned with manufacturing, where production capability is distributed across machines, software, identities, materials, quality checks, and external partners. Supply-chain ransomware studies also show that an attack on one organization can affect production, suppliers, customers, and logistics [20]--[22]. For manufacturing recovery, the relevant object is therefore not a single device or database but a dependency network that must support a minimum viable production function.

This paper builds on but differs from general resilience work. General cyber-resilience frameworks often describe phases such as prepare, absorb, recover, and adapt. Those phases are useful, but they do not specify what “recover” means for a plant that must coordinate identity, MES, historian, quality release, OT reconnection, material flow, and supplier communication. MVF Recovery operationalizes this idea by asking which constrained production mission can be resumed under current trust and evidence conditions.

\subsection{Forensics, evidence, and proof-oriented recovery}

Digital forensics contributes a second missing ingredient: evidence. Factory operators need to know not only that a backup exists, but whether it is clean enough to restore; not only that an identity service is running, but whether credentials and privileges can be trusted; not only that OT equipment is reachable, but whether reconnection is safe. Forensic case studies of ransomware in factory and OT settings show the value of network diagrams, event logs, digital evidence, timelines, and attack-framework mappings [18], [31]. Forensic readiness research argues that evidence should be proactively collected so that investigations and recovery can proceed more effectively [17]. These works motivate treating proof-of-recovery as a practical evidence problem, not merely a legal or post-incident reporting issue.

The term proof-of-recovery is used cautiously in this paper. It does not mean formal proof in the mathematical sense, nor does it mean formal safety certification. It means an auditable evidence bundle that justifies a recovery decision: restore source, compromise assessment, identity state, configuration validation, dependency consistency, OT reintegration check, degraded-mode limits, monitoring plan, and residual risk. This is the kind of evidence that allows responders, plant managers, safety stakeholders, and auditors to understand why a constrained production restart is defensible.

\subsection{Recovery exercises, models, and preparedness}

Several studies address incident response modelling, cyber-resilience exercises, and OT preparedness. Hybrid exercises combining simulation and tabletop activities help participants reason about cyber incidents affecting plants and identify safety responses [32]. Incident response modelling languages support representation of missions, resources, vulnerabilities, detection, mitigation, and recovery mechanisms [33]. OT preparedness research emphasizes risk-based safety scoping for security tests and recovery exercises in ICS/OT environments [23], [24], [34]. These works support the need for structured recovery planning but do not by themselves define a manufacturing-specific recovery objective after ransomware.

IJCIP has published work on ICS attack lessons and cyber incident response/recovery frameworks for operators [23], [24]. This paper complements that work by narrowing the lens to ransomware recovery in critical manufacturing and by synthesizing a recovery-failure taxonomy with a capability objective. It is therefore not a replacement for incident response standards, business continuity planning, or OT safety engineering. It is a bridge between those practices and the specific recovery question that ransomware creates: when can a plant resume a constrained but valid production mission?

\subsection{Positioning of this paper}

This paper sits at the intersection of these strands. It does not propose another ransomware detector, backup mechanism, forensic tool, or general incident-response checklist. Instead, it asks how manufacturing recovery should be conceptualized when production depends on interdependent IT/OT, physical, organizational, and supply-chain systems. The contribution is a synthesis: an evidence-backed taxonomy of recovery failure modes and a capability-centric recovery objective, MVF Recovery, that connects dependency modelling, trust reconstruction, evidence, OT safety, degraded operation, and supplier coordination.

The broader review corpus also includes CPS cyber-resilience, threat-modelling, forensic-readiness, and dependency-analysis studies [17], [35]--[39], organizational recovery modelling [40], primary incident or disruption updates [41]--[44], backup guidance [45], recent industrial threat landscape reports and recovery guidance [46]--[52], industrial security standards and review-method guidance [53]--[65], and additional manufacturing supply-chain or PLC-vulnerability work [66], [67]. These sources inform background, method design, or candidate-gap interpretation; they are not treated as high-impact taxonomy anchors unless full-text or source-page evidence supports the stronger claim.

%% file: sections/03_methodology.tex
\section{Methodology}

\subsection{Review design and scope}

We conducted a PRISMA-guided multivocal review of ransomware recovery in critical manufacturing infrastructure. The review combines academic literature, government and standards guidance, threat frameworks, industrial reports, public incident material, and company or regulatory disclosures. The goal was not to estimate population prevalence of every recovery failure. Instead, the goal was to identify and calibrate source-supported recovery failure modes and derive a defensible recovery objective for manufacturing systems.

The review is multivocal because relevant evidence is distributed across peer-reviewed papers, standards, government guidance, threat frameworks, incident reports, company statements, and public disclosures. Manufacturing ransomware recovery is rarely documented in a single academic literature stream. Academic work is stronger on mechanisms, models, and terminology; government and standards sources are stronger on recovery guidance; company and incident sources are stronger on operational disruption; and forensic or OT case studies provide detailed technical anchors. The synthesis therefore treats source type as part of claim calibration rather than as a reason to exclude non-academic material.

\subsection{Research questions and synthesis logic}

The four research questions introduced in Section 1 structured the synthesis. RQ1 guided source coding for failure modes. RQ2 guided cross-mode interpretation: whether evidence supported the argument that asset restoration is insufficient. RQ3 guided derivation of the MVF Recovery objective. RQ4 guided extraction of lifecycle steps and benchmarking dimensions. This sequence is important: MVF Recovery is not introduced as an independent design idea, but as a response to the observed failure modes.

\subsection{Information sources and search coverage}

The academic search covered Google Scholar via Publish or Perish, PubMed, IEEE Xplore, ACM Digital Library, ScienceDirect, and Scopus. Web of Science was recorded as unavailable because institutional access was not available. Grey and practitioner sources included CISA, NIST, NCSC, ENISA, MITRE ATT\&CK for ICS, Kaspersky ICS CERT, Dragos, public company statements, SEC filings, and public incident reports. Search strings combined ransomware terms with manufacturing, ICS, OT, cyber-physical systems, SCADA, recovery, restoration, resilience, backup, incident response, and supply-chain terms. The Google Scholar/Publish or Perish queries were: \emph{ransomware manufacturing recovery}; \emph{ransomware industrial control system recovery}; \emph{ransomware operational technology incident response}; \emph{ransomware backup restore forensic verification}; and \emph{ransomware cyber resilience manufacturing}. ScienceDirect exports were limited to the first 100 records per query, sorted by ScienceDirect relevance, because the interface did not provide a usable export-all option. The ScienceDirect component therefore contributes 500 exported records from five capped queries, while total hit counts are retained only as search-scope context.

The academic workflow produced 2,307 exported records. After deduplication, 1,987 unique academic records remained. Title/abstract screening was applied to these academic records and excluded 1,152 records, leaving 835 records for source-level eligibility. Source-level eligibility retained 342 eligible core academic records and 427 background academic records, while 66 academic records were excluded at eligibility. Separately, 28 grey-literature and incident sources were added after academic screening and handled through source-page verification rather than title/abstract screening. The final multivocal corpus available for synthesis therefore comprised 797 retained records or sources: 342 eligible core academic records, 427 background academic records, and 28 grey or incident sources. The main article cites 76 sources selected as high-impact, representative, methodological, guidance, or incident anchors. The difference between the 342 eligible core records and the 76 cited sources reflects journal-length citation selection and evidence-role prioritisation; it should not be read as a claim that all 342 eligible records are individually cited, page-level verified, or used as one-to-one support for every taxonomy claim.

Table~\ref{tab:prisma-flow} records the manuscript-level flow counts used for screening and eligibility.

{\small
\begin{longtable}{@{}>{\RaggedRight\arraybackslash}p{0.44\textwidth} >{\RaggedRight\arraybackslash}p{0.18\textwidth} >{\RaggedRight\arraybackslash}p{0.32\textwidth}@{}}
\caption{PRISMA-guided flow counts for the multivocal review.}\label{tab:prisma-flow}\\
\toprule
Stage & Count & Note \\
\midrule
\endfirsthead
\toprule
Stage & Count & Note \\
\midrule
\endhead
Exported academic records & 2,307 & Database and academic-search exports before deduplication. \\
Unique academic records after deduplication & 1,987 & Duplicate records removed. \\
Academic records excluded at title/abstract screening & 1,152 & First-stage screening applied only to academic records. \\
Academic records entering source-level eligibility & 835 & 1,987 minus 1,152. \\
Eligible core academic records retained & 342 & Records eligible for direct synthesis roles; not all are cited in the main article. \\
Background academic records retained & 427 & Records relevant to context, definitions, or adjacent concepts. \\
Academic records excluded at eligibility & 66 & 342 + 427 + 66 = 835. \\
Grey-literature and incident sources added after screening & 28 & Government, standards, company, regulatory, and incident sources handled by source-page verification. \\
Final multivocal corpus available for synthesis & 797 & 342 eligible core academic records + 427 background academic records + 28 grey/incident sources. \\
\bottomrule
\end{longtable}
}

\subsection{Review workflow}

\textbf{Box 1 summarizes the review and evidence-calibration workflow.} Detailed database-level search and export-count logs, screening decisions, representative full-text/source-page verification anchors, methodological consistency checks, the adjudicated failure-mode evidence summary, and a cited-source evidence matrix are provided as supplementary material. The supplementary upload contains S1 search strategy and source log, S2 PRISMA flow counts, S3 aggregate screening decisions, S4 representative full-text/source-page verification anchors, S5 adjudicated failure-mode evidence summary, S6 methodological consistency check, S7 candidate research gaps, and S8 cited-source evidence matrix.

\input{box_methodology_workflow_v2_11i.tex}

This workflow is PRISMA-guided rather than a claim of full PRISMA compliance. The review includes transparent identification, screening, eligibility, and inclusion counts, but access restrictions and heterogeneous source types prevent a claim that every retained record was fully read at page level.

\subsection{Screening and source-level eligibility}

Screening proceeded in two stages. First, title/abstract screening removed records unrelated to ransomware, recovery, manufacturing, ICS/OT, CPS, critical infrastructure, or cyber resilience. Second, source-level eligibility assessed whether the source contributed to one of four roles: core evidence, background evidence, methodological support, or exclusion. Core evidence included sources that directly supported recovery failure modes, MVF assumptions, recovery lifecycle steps, or manufacturing/critical-infrastructure recovery constraints. Background evidence included sources relevant to general ransomware, ICS security, cyber resilience, or forensics but not directly tied to the taxonomy.

For access-restricted records where full text was unavailable, eligibility was based on title, abstract, venue, DOI metadata, and available source snippets. Such records were not used as direct support for high-impact taxonomy claims unless full-text or source-page evidence was later verified. Records that could not be retrieved were retained only as metadata-level or background evidence.

\subsection{Source quality and permitted claim strength}

Because the review combines academic and grey literature, source quality was used to calibrate the strongest claim a source could support.

\input{tables/source_quality_rubric_v2_4.tex}

This rubric prevents a public statement or news report from being treated as equivalent to a full forensic case study. It also allows official guidance to support recovery-practice claims while avoiding incident-frequency claims.

\subsection{Full-text/source-page verification}

High-impact taxonomy claims were verified using accessible full text, official guidance pages, company statements, standards documents, public incident materials, and author-supplied PDFs. Verification focused on whether the source explicitly or inferentially supported a claimed recovery failure mode. Examples include factory ICS forensic cases with event logs and timelines [18], OT ransomware analyses [31], supply-chain recovery cases [21], critical-infrastructure cyber-resilience models [16], and ransomware in food or energy supply chains [15], [22]. Metadata-only records were not treated as direct support for strong claims.

Verification did not mean that every one of the 342 eligible core academic records was read as a complete full text. Instead, full-text/source-page verification was performed for high-impact claims and representative anchors. This distinction is necessary because some records were paywalled or only available as metadata at the time of review. The manuscript therefore uses cautious language: source-level eligibility for the full corpus, full-text/source-page verification for high-impact claims.

\subsection{Coding validation and adjudication}

The coding scheme assigned each source to one or more failure modes and evidence categories: direct support, inferred support, background support, or exclusion. Because one source can discuss multiple recovery issues, a source may support more than one failure mode. Evidence counts therefore represent adjudicated source-level support records rather than incident frequency.

Because the final manuscript is single-authored, the review does not claim full independent inter-rater reliability for the corpus. Instead, a structured consistency check was performed across stratified validation categories covering title/abstract screening, source-level eligibility, and failure-mode support spot-checking. Ambiguous cases were retained for adjudication and resolved conservatively: disputed direct support was downgraded unless explicit full-text/source-page evidence supported the stronger classification. A failure mode was retained in the main taxonomy only when it had at least one verified high-impact anchor and cross-source conceptual support; otherwise, it was treated as contextual support or moved to the candidate research gaps. This approach is reported transparently in the supplementary material and is intended as claim calibration rather than as a population-level reliability estimate.

\subsection{Claim-strength calibration}

The synthesis distinguishes three levels of claim. \textbf{Evidence-backed failure modes} are supported by verified sources and appear in the main taxonomy. \textbf{Inferred or contextual mechanisms} are discussed only where evidence indicates a plausible relation but does not directly document repeated recovery failure. \textbf{Candidate research gaps} are analytically important but under-reported in public evidence and are therefore moved to Section 8 rather than included in the main taxonomy.

We did not conduct bibliometric analysis. The purpose of this paper is not to map citation networks, author communities, or the intellectual structure of a field. The purpose is to synthesize recovery failure mechanisms and derive an operational recovery objective. Bibliometric analysis would be useful for a different paper on the development of cyber-resilience or ICS ransomware research, but it would not by itself answer the recovery-decision questions addressed here.

\subsection{Supplementary material strategy}

For journal submission, the main text reports only the audit trail needed to evaluate the synthesis. The supplementary material provides aggregate search strategy records, PRISMA-style flow counts, aggregate screening decisions, representative full-text/source-page verification anchors, an adjudicated failure-mode evidence summary, methodological consistency checks, a cited-source evidence matrix, and candidate research gaps. This separation is important because the review is evidence-heavy, but a journal article should not reproduce every coded record in the main body. The supplement allows reviewers to inspect the provenance of representative claims while keeping the article focused on the taxonomy, MVF objective, lifecycle, and research agenda.

\subsection{Threats to methodological validity}

The methodology is limited by public reporting bias, access restrictions, and the sensitivity of manufacturing recovery data. Large public companies and high-profile incidents are more visible than smaller firms. Some incidents are described only through business updates rather than technical recovery reports. The conservative mitigation is to avoid prevalence claims, avoid using unretrieved records as direct support, maintain a source-quality rubric, and separate under-reported candidate patterns from evidence-backed failure modes.

%% file: box_methodology_workflow_v2_11i.tex
\begin{center}
\fbox{\begin{minipage}{0.92\linewidth}\small
\textbf{Academic stream:} academic database search $\rightarrow$ deduplication $\rightarrow$ title/abstract screening $\rightarrow$ source-level eligibility $\rightarrow$ high-impact full-text/source-page verification.\\[0.35em]
\textbf{Grey/incident stream:} grey and incident source collection $\rightarrow$ source-page verification.\\[0.35em]
\textbf{Synthesis stream:} evidence classification $\rightarrow$ structured consistency checking and conservative adjudication $\rightarrow$ evidence-backed taxonomy synthesis $\rightarrow$ MVF Recovery derivation and benchmarking directions.
\end{minipage}}
\par\vspace{0.35em}
{\small\textbf{Box 1.} PRISMA-guided multivocal review and evidence-calibration workflow.}
\end{center}

%% file: tables/source_quality_rubric_v2_4.tex
{\small
\begin{longtable}{@{}>{\RaggedRight\arraybackslash}p{0.23\textwidth} >{\RaggedRight\arraybackslash}p{0.35\textwidth} >{\RaggedRight\arraybackslash}p{0.36\textwidth}@{}}
\caption{Source-quality rubric and permitted claim strength.}\label{tab:source-quality}\\
\toprule
Source type & Typical use in synthesis & Maximum claim strength without corroboration \\
\midrule
\endfirsthead
\toprule
Source type & Typical use in synthesis & Maximum claim strength without corroboration \\
\midrule
\endhead
Peer-reviewed academic paper & Mechanisms, models, experiments, case analysis, conceptual framing & Direct or inferred support, depending on content. \\
Government or standards guidance & Recommended recovery, backup, identity, logging, OT, and resilience practices & Direct support for recommended practices and recovery constraints. \\
Threat framework or industrial threat report & Tactics, techniques, OT impact categories, sector threat context & Direct support for attack/recovery mechanism if source is specific; otherwise background. \\
Company statement or SEC filing & Incident facts, business disruption, staged restart, operational impact & Direct support for reported incident facts; inferred support for mechanisms only when corroborated. \\
Public news report & Timeline context, public incident discovery, sector impact & Background or inferred support unless triangulated with primary sources. \\
Vendor blog or practitioner article & Threat context, technical observations, practitioner framing & Background unless corroborated by academic, government, or primary incident material. \\
\bottomrule
\end{longtable}
}

%% file: sections/04_evidence_backed_recovery_failure_taxonomy.tex
\section{Evidence-Backed Recovery Failure Taxonomy}

\subsection{Overview and interpretation of evidence counts}

This section answers RQ1 by presenting nine evidence-backed recovery failure modes. The taxonomy is organized around four recovery problem classes: dependency failures, trust and verification failures, reintegration failures, and operational capability failures. The taxonomy is not a prevalence estimate. Evidence counts represent adjudicated source-level support records rather than incident frequency. A single source may support multiple failure modes if it discusses multiple recovery issues.

\input{tables/taxonomy_classes_v2_5.tex}

The four candidate patterns removed from the main taxonomy are discussed in Section 8: asset-centric recovery metrics, manual playbook brittleness, reinfection during recovery, and quality-release blocking. Their weak public support is treated as a reporting and research gap rather than as confirmed recurring evidence.

\subsection{Representative evidence anchors}

Table~\ref{tab:evidence-anchors} gives representative anchors for the main taxonomy. It is not a complete coding matrix; it is a compact guide showing how the strongest claims are supported.

\input{tables/representative_evidence_anchors_v2_4.tex}

\subsection{FM01: Dependency blindness}

\textbf{Definition.} Dependency blindness occurs when recovery planning treats systems as independent assets rather than as components of a production mission.

\textbf{Recovery mechanism.} Manufacturing capability depends on chains of digital, physical, human, and external dependencies. MES may require identity, order data, recipes, quality records, historian data, and network access. Production may require supplier confirmation, logistics systems, packaging lines, and manual sign-off. A single missing dependency can prevent a restored application from producing valid output.

\textbf{Evidence anchor.} Supply-chain and critical-infrastructure studies describe interdependent process components and cascading disruption after cyber incidents [16], [20], [21]. ICS/IIoT ransomware studies show that edge gateways, supervisory workstations, and control-system components can bridge IT and OT dependencies [13], [14], [18].

\textbf{MVF implication.} MVF Recovery must begin with a dependency graph. The recovery target is not a list of machines but a set of dependencies sufficient to support a constrained production mission.

\subsection{FM02: Untrusted restore point and backup over-trust}

\textbf{Definition.} Backup over-trust occurs when responders assume that the newest or most available backup is clean, complete, and operationally suitable without verifying compromise state, dependency consistency, or attacker persistence.

\textbf{Recovery mechanism.} Ransomware can delete, encrypt, corrupt, or pre-compromise backups. Even when backups are intact, restoring a contaminated identity system, engineering workstation, or configuration repository can reintroduce compromise. Guidance from CISA and NCSC emphasizes offline, tested, and scanned backups and clean recovery environments [4], [6]. Ransomware recovery research also shows why backup-only restoration can be insufficient [25], [26].

\textbf{Evidence anchor.} Storage recovery studies document backup spoliation and limitations of whole-disk restoration [25], [26]. Ransomware-readiness and forensic-readiness sources reinforce the need to select clean states using evidence rather than restore recency alone [4]--[6], [17].

\textbf{MVF implication.} MVF planning must choose restore points according to cleanliness, dependency consistency, and evidence quality, not only recency.

\subsection{FM03: Identity trust collapse}

\textbf{Definition.} Identity trust collapse occurs when authentication, authorization, privileged accounts, domain controllers, tokens, service accounts, remote access, or administrative workstations can no longer be trusted.

\textbf{Recovery mechanism.} Manufacturing recovery depends on authenticated operators, engineers, administrators, service accounts, and vendor connections. If identity infrastructure is compromised, restored assets may be reachable but unsafe to use. Ransomware often spreads through credential theft, lateral movement, remote administration tools, and shared services [18], [31].

\textbf{Evidence anchor.} Factory ICS forensic cases describe initial access through remote services, local administrator misuse, credential theft, and lateral movement [18]. Industrial virtual-lab and ICS vulnerability studies show how misconfiguration, weak authentication, and exposed control-system dependencies can cascade into ransomware impact [28], [68]. Supply-chain recovery cases also show that communication, authorization, and coordination capacity are part of recovery [21].

\textbf{MVF implication.} MVF Recovery requires an identity-clean room: validated administrator accounts, reset credentials, trusted privileged workstations, and controlled access paths before production systems are restored.

\subsection{FM04: No proof-of-recovery}

\textbf{Definition.} No proof-of-recovery occurs when responders cannot demonstrate that recovered systems are clean, configured correctly, operationally valid, and safe to use.

\textbf{Recovery mechanism.} Restoration without evidence creates uncertainty. A rebuilt server may be online, but responders may not know whether malware persists, whether logs support the recovery decision, whether the backup was clean, whether credentials remain compromised, or whether OT reconnection is safe. This is a practical evidence problem rather than a formal theorem-proving problem.

\textbf{Evidence anchor.} Digital forensic readiness research argues for proactive evidence collection before and during ransomware incidents [17]. Factory ransomware forensic studies show the value of network diagrams, event logs, digital evidence, and timelines [18]. Incident investigations such as LockerGoga/Norsk Hydro also illustrate the volume and complexity of evidence required for recovery decisions [69], [70].

\textbf{MVF implication.} MVF Recovery must include an evidence bundle: source of restore point, compromise assessment, credential state, configuration validation, test results, and monitoring plan.

\subsection{FM05: Unsafe OT reconnection}

\textbf{Definition.} Unsafe OT reconnection occurs when recovered IT or OT assets are reconnected to production networks before their safety, trust, configuration, and process impact have been validated.

\textbf{Recovery mechanism.} OT systems interact with physical processes. Reconnecting an HMI, engineering workstation, historian, or PLC-facing system can affect commands, visibility, alarms, setpoints, and operator decisions. Energy-delivery and ICS ransomware analyses show that loss of situational awareness or command pathways can be operationally serious even without direct physical destruction [15], [29], [30].

\textbf{Evidence anchor.} ICS ransomware studies demonstrate ransomware effects on PLCs, supervisory systems, and edge gateways [13], [14]. OT preparedness research emphasizes safety-risk scoping and operational constraints in response and recovery [34].

\textbf{MVF implication.} MVF Recovery requires staged reintegration gates: offline validation, isolated recovery networks, safety checks, operator approval, limited-scope reconnection, and monitored production resumption.

\subsection{FM06: Segmentation assumption failure}

\textbf{Definition.} Segmentation assumption failure occurs when recovery assumes that IT/OT segmentation, air gaps, or network zoning prevented propagation, even though real paths exist through remote access, domain trust, shared services, removable media, gateways, or vendor connections.

\textbf{Recovery mechanism.} Segmentation is often partial, degraded, misconfigured, or bypassed during operations and maintenance. Brownfield IIoT gateways and remote access can bridge physical and cyber domains [14]. Vulnerability studies identify weak authentication, improper network configuration, poor logging, and outdated software as common ICS weaknesses [68].

\textbf{Evidence anchor.} Factory forensic cases and virtual-lab studies show lateral movement across network paths and the role of AD, remote access, and segmentation flaws [18], [28]. ICS ransomware work demonstrates that physical isolation alone is not sufficient to reason about recovery [13].

\textbf{MVF implication.} Recovery planning must verify actual communication paths, not assumed architecture diagrams. Segmentation evidence should be treated as a recovery input.

\subsection{FM07: Capability mismatch}

\textbf{Definition.} Capability mismatch occurs when restored assets do not correspond to restored production capability. A factory may restore servers but remain unable to produce, inspect, release, ship, or coordinate work.

\textbf{Recovery mechanism.} Production requires end-to-end capability: order intake, material availability, scheduling, process control, quality validation, packaging, logistics, and customer communication. Partial IT recovery can leave factories functionally unrecovered. Conversely, limited production may resume without full enterprise restoration if the necessary capability dependencies are satisfied.

\textbf{Evidence anchor.} Supply-chain recovery studies, factory forensic cases, and critical-infrastructure resilience models all support the distinction between restored components and restored process continuity [16], [18], [20]--[22]. Public incidents involving manufacturing and supply-chain disruption show staged restart, constrained operation, and downstream effects [69], [71]--[76].

\textbf{MVF implication.} MVF Recovery measures recovery by capability outcomes: what product can be made, at what volume and quality, using which trusted systems and procedures.

\subsection{FM08: Degraded mode unmanaged}

\textbf{Definition.} Degraded mode unmanaged occurs when manual or partial operations are used without defined limits, safety controls, evidence requirements, or transition criteria.

\textbf{Recovery mechanism.} Degraded operation can preserve continuity but can also create undocumented workarounds, quality risk, safety exposure, and inconsistent recovery evidence. Manufacturing incidents and resilience exercises show that organizations may rely on manual operations or constrained processes during cyber disruption [21], [32], [69].

\textbf{Evidence anchor.} Hybrid OT exercises emphasize the need to understand plant safety responses and operational collaboration [32]. Supply-chain recovery case evidence shows the importance of structured readiness, response, recovery, and growth phases [21]. OT preparedness research highlights safety-risk scoping for recovery and testing activities [34].

\textbf{MVF implication.} MVF Recovery can include degraded production, but only if constraints, authority, safety checks, quality rules, and exit criteria are explicit.

\subsection{FM09: Supplier dependency failure}

\textbf{Definition.} Supplier dependency failure occurs when recovery planning focuses on internal assets while overlooking suppliers, customers, logistics providers, remote vendors, outsourced IT/OT services, or shared platforms.

\textbf{Recovery mechanism.} Manufacturing recovery often depends on external parties: raw-material suppliers, contract manufacturers, equipment vendors, logistics providers, cloud systems, customers, and regulators. A supplier incident can halt a factory, while a factory incident can propagate to customers and partners.

\textbf{Evidence anchor.} Integrated supply-chain modelling shows how ransomware at one firm can affect dependent firms [20]. A digital supply-chain recovery case documents recovery through both intra-firm and inter-firm capabilities [21]. Food supply-chain ransomware work describes how digitalization and integration increase vulnerability and how disruption can affect production, logistics, and food security [22].

\textbf{MVF implication.} MVF Recovery must include external dependencies and communication channels. A plant is not minimally viable if it cannot receive required inputs, ship validated outputs, or coordinate with critical partners.

%% file: tables/taxonomy_classes_v2_5.tex
{\small
\begin{longtable}{@{}>{\RaggedRight\arraybackslash}p{0.23\textwidth} >{\RaggedRight\arraybackslash}p{0.38\textwidth} >{\RaggedRight\arraybackslash}p{0.33\textwidth}@{}}
\caption{Recovery failure-mode classes and core recovery questions.}\label{tab:taxonomy-classes}\\
\toprule
Class & Failure modes & Core recovery question \\
\midrule
\endfirsthead
\toprule
Class & Failure modes & Core recovery question \\
\midrule
\endhead
Dependency failures & FM01 Dependency blindness & What must exist together for production to resume? \\
Trust and verification failures & FM02 Backup over-trust; FM03 Identity trust collapse; FM04 No proof-of-recovery & Which restored systems, identities, data, and configurations can be trusted? \\
Reintegration failures & FM05 Unsafe OT reconnection; FM06 Segmentation assumption failure & When and how can recovered systems reconnect to production safely? \\
Operational capability failures & FM07 Capability mismatch; FM08 Degraded mode unmanaged; FM09 Supplier dependency failure & What constrained production capability can be resumed, and under what limits? \\
\bottomrule
\end{longtable}
}

%% file: tables/representative_evidence_anchors_v2_4.tex
{\footnotesize
\begin{longtable}{@{}>{\RaggedRight\arraybackslash}p{0.22\textwidth} >{\RaggedRight\arraybackslash}p{0.35\textwidth} >{\RaggedRight\arraybackslash}p{0.37\textwidth}@{}}
\caption{Representative evidence anchors for the recovery failure taxonomy.}\label{tab:evidence-anchors}\\
\toprule
Failure mode & Representative anchors & What the anchors support \\
\midrule
\endfirsthead
\toprule
Failure mode & Representative anchors & What the anchors support \\
\midrule
\endhead
FM01 Dependency blindness & Integrated supply-chain ransomware modelling [20]; critical-infrastructure process-continuity model [16]; IIoT edge-gateway ransomware analysis [14] & Recovery depends on interdependent process components, not isolated assets. \\
FM02 Backup over-trust & Storage/restore limitation studies [25], [26]; ransomware and forensic-readiness guidance [4]--[6], [17] & Clean restore points must be verified, not assumed from recency or backup existence. \\
FM03 Identity trust collapse & Factory forensic case [18]; LockBit OT analysis [31]; industrial virtual lab with AD misconfiguration [28] & Recovery can fail if credentials, privileged access, remote administration, or domain trust remain compromised. \\
FM04 No proof-of-recovery & Factory forensic timeline and evidence [18]; forensic readiness framework [17]; LockerGoga/Norsk Hydro evidence complexity [69], [70] & Recovery decisions require evidence bundles, not only operational intuition. \\
FM05 Unsafe OT reconnection & ICS ransomware realism study [13]; OT preparedness thesis [34]; energy-delivery ransomware analysis [15] & OT reconnection must be staged, safety-aware, and monitored. \\
FM06 Segmentation assumption failure & Brownfield IIoT edge-gateway analysis [14]; ICS vulnerability assessment [68]; factory forensic case [18] & Assumed segmentation, air gaps, and diagrams may not reflect actual recovery paths. \\
FM07 Capability mismatch & Supply-chain recovery case [21]; critical-infrastructure cyber-resilience model [16]; public manufacturing incidents [71]--[76] & Restored systems do not automatically equal restored production capability. \\
FM08 Degraded mode unmanaged & Hybrid cyber-resilience exercise [32]; supply-chain recovery case [21]; OT preparedness work [34] & Manual or degraded operation requires explicit limits, safety checks, and exit criteria. \\
FM09 Supplier dependency failure & Supply-chain ransomware economics [20]; digital supply-chain recovery case [21]; food supply-chain ransomware analysis [22] & External partners and logistics can determine whether a factory is operationally recovered. \\
\bottomrule
\end{longtable}
}

%% file: sections/05_minimum_viable_factory_recovery.tex
\section{Minimum Viable Factory Recovery}

\subsection{Motivation}

Section 4 shows that manufacturing ransomware recovery cannot be judged only by asset availability. The nine failure modes converge on one problem: responders need an objective that represents constrained production capability under uncertainty. MVF Recovery provides that objective. It asks which production mission can be resumed safely, with trusted identities and systems, with sufficient evidence, and within explicit operational limits.

MVF Recovery should not be confused with full recovery. A factory may reach MVF while still operating below normal throughput, using manual workarounds, restricted product families, limited supplier interfaces, enhanced monitoring, or temporary quality controls. Nor should MVF be interpreted as formal safety certification. It is an analytical recovery objective that helps responders make and justify staged restart decisions.

\subsection{Definition}

\textbf{Minimum Viable Factory Recovery} is the smallest set of trusted systems, identities, data, network paths, OT interfaces, procedures, people, and external dependencies required to resume a constrained but valid production mission after ransomware disruption.

An MVF mission should specify at least the elements in Table~\ref{tab:mvf-elements}.

{\small
\begin{longtable}{@{}>{\RaggedRight\arraybackslash}p{0.27\textwidth} >{\RaggedRight\arraybackslash}p{0.67\textwidth}@{}}
\caption{Minimum elements of an MVF mission statement.}\label{tab:mvf-elements}\\
\toprule
Element & Example question \\
\midrule
\endfirsthead
\toprule
Element & Example question \\
\midrule
\endhead
Production scope & Which product family, line, cell, or batch can be produced? \\
Throughput and duration & At what volume and for how long? \\
Safety envelope & Which safety checks and operating constraints apply? \\
Quality validity & Which inspection and release requirements must hold? \\
Dependency set & Which IT, OT, identity, data, supplier, and logistics dependencies are required? \\
Trust state & Which identities, systems, backups, and network paths are trusted? \\
Evidence bundle & What evidence supports the restart decision? \\
Degraded-mode limits & What manual workarounds are allowed, and when do they expire? \\
Monitoring and rollback & What signals trigger escalation, rollback, or expansion? \\
\bottomrule
\end{longtable}
}

\subsection{Distinction from adjacent recovery concepts}

MVF Recovery is related to, but distinct from, business continuity planning, disaster recovery, mission-essential function analysis, and RTO/RPO targets. Those concepts remain necessary, but they do not by themselves answer the ransomware-specific restart question: which constrained production mission is safe, trusted, evidence-supported, and operationally meaningful under the current compromise state? Table~\ref{tab:mvf-vs-adjacent} summarizes the distinction.

{\footnotesize
\begin{longtable}{@{}>{\RaggedRight\arraybackslash}p{0.18\textwidth} >{\RaggedRight\arraybackslash}p{0.24\textwidth} >{\RaggedRight\arraybackslash}p{0.27\textwidth} >{\RaggedRight\arraybackslash}p{0.23\textwidth}@{}}
\caption{Distinguishing MVF Recovery from adjacent recovery concepts.}\label{tab:mvf-vs-adjacent}\\
\toprule
Concept & Usual focus & Ransomware-specific gap & MVF Recovery addition \\
\midrule
\endfirsthead
\toprule
Concept & Usual focus & Ransomware-specific gap & MVF Recovery addition \\
\midrule
\endhead
Business continuity planning / BIA & Prioritizing business processes, resources, and continuity arrangements before disruption [62]. & May assume known dependencies and usable recovery inputs. It does not necessarily validate trust state after compromise. & Instantiates a current, limited production mission using evidence about dependencies, trust, identity, data, and suppliers. \\
Disaster recovery & Restoring systems, services, and data to operating states [62], [65]. & Can optimize asset restoration without proving backup cleanliness, identity integrity, OT reconnection safety, or production validity. & Evaluates whether restored components form a safe and valid manufacturing capability. \\
Mission-essential function / MBCO-style target & Maintaining the smallest business or mission function needed for continuity. & Often remains at process level and may not specify cyber-physical dependencies, restore cleanliness, or OT evidence gates. & Translates the mission into required systems, credentials, data, OT interfaces, people, and external dependencies. \\
RTO / RPO & Setting time-to-recover and data-loss thresholds. & Fast or recent restoration may be unsafe if credentials, backups, or engineering workstations are compromised. & Allows a slower or older restore point when it is better evidenced and sufficient for a constrained mission. \\
Incident response and OT recovery & Containment, eradication, recovery coordination, and industrial response practices [24], [27]. & Does not by itself define which constrained production mission should resume first. & Provides a capability-centric decision objective for staged restart after containment and verification. \\
\bottomrule
\end{longtable}
}

This distinction is central to the contribution. MVF Recovery is not a replacement name for BCP, DR, MBCO, RTO, or RPO. It is a ransomware-recovery decision construct that adds trust state, identity collapse, restore cleanliness, OT reconnection evidence, quality validity, and supplier dependency to the definition of a restartable factory capability.

\subsection{Analytical model}

Let a manufacturing environment be represented as a dependency graph \textbf{G = (V, E)}, where nodes are systems, identities, data stores, OT components, procedures, people, materials, and external partners, and edges represent dependency relations required for production. Let \textbf{M} be a candidate production mission, such as producing a restricted product family on one line at reduced throughput. Let \textbf{D(M)} be the subset of nodes and edges required for that mission.

Each node or edge has a recovery state with at least four attributes:

\begin{itemize}[leftmargin=*]
\item \textbf{Availability:} whether the component is operationally reachable or usable.
\item \textbf{Trust:} whether the component is believed clean enough for the mission.
\item \textbf{Evidence:} whether the trust and configuration claims are supported by logs, forensic findings, tests, or source verification.
\item \textbf{Safety/operational status:} whether use of the component is allowed under current plant procedures.
\end{itemize}

A compact feasibility condition is:
\[
\mathrm{MVF}(M) \iff \forall d \in D(M): A(d) \geq a_{\min} \land T(d) \geq t_{\min} \land E(d) \geq e_{\min} \land S(d)=\mathrm{approved}.
\]
Here, \(A\), \(T\), \(E\), and \(S\) denote availability, trust, evidence sufficiency, and safety/operational approval. A mission reaches MVF when all critical dependencies in \(D(M)\) satisfy these thresholds and when degraded-mode limits and rollback conditions are defined. This is deliberately a decision model, not a proof system. Its value is to make recovery assumptions explicit.

\subsection{MVF success conditions}

An MVF decision should satisfy five conditions.

\begin{enumerate}[leftmargin=*]
\item \textbf{Capability condition:} the selected mission can produce a defined output at a defined quality and throughput level.
\item \textbf{Dependency condition:} all required dependencies for that mission are available or replaced by approved degraded procedures.
\item \textbf{Trust condition:} identities, restore points, administrative workstations, network paths, and OT interfaces used by the mission are trusted enough for constrained operation.
\item \textbf{Evidence condition:} recovery decisions are supported by an evidence bundle that can be reviewed.
\item \textbf{Reintegration condition:} OT and supplier/customer connections are reintroduced through controlled gates and monitored after restart.
\end{enumerate}

These conditions link directly to the failure modes. Dependency blindness violates the dependency condition. Backup over-trust and identity collapse violate the trust condition. No proof-of-recovery violates the evidence condition. Unsafe OT reconnection and segmentation assumption failure violate the reintegration condition. Capability mismatch, unmanaged degraded mode, and supplier dependency failure violate the capability and dependency conditions.

\subsection{Taxonomy-to-MVF mapping}

\input{tables/taxonomy_to_mvf_mapping_v2_4.tex}

\subsection{Illustrative MVF decision example}

Consider a manufacturer whose enterprise domain, MES, historian, and supplier-order interface are disrupted. Full recovery would require rebuilding the enterprise identity environment, restoring MES, validating historian integrity, restoring ERP integration, reconnecting engineering workstations, and confirming supplier and customer data flows. MVF Recovery does not ask whether all of these systems are back. It asks whether a narrower mission can be resumed defensibly. For example, the plant may select one product family on one production cell for a 48-hour constrained restart. That mission may require a clean administrator workstation, reset operator credentials, a verified MES restore for the selected product family, offline recipe validation, a manually approved quality-release procedure, an isolated OT reconnection path, and manual supplier confirmation.

This example shows why MVF is not simply “partial recovery.” Partial recovery can be accidental: some systems happen to be online, and production resumes around them. MVF is deliberate: the organization defines the mission, dependency set, evidence requirements, degraded-mode limits, and monitoring conditions before restart. If the selected MES backup is recent but suspected contaminated, it may be rejected in favour of an older verified restore point. If identity cannot be trusted, the mission may be delayed even if production applications are reachable. If supplier communication is unavailable, the mission may be restricted to products with confirmed material availability. The same plant could therefore choose different MVF missions at different times during the incident as evidence improves.

\subsection{Relationship to business continuity and incident response}

MVF Recovery complements, rather than replaces, business continuity planning, disaster recovery, incident response, and OT safety engineering. Business continuity planning often defines priority processes and continuity targets. Incident response contains and eradicates adversary activity. Disaster recovery restores systems. OT safety engineering governs plant safety. MVF sits between these practices and asks whether the restored components are sufficient for a specific constrained production mission.

This positioning is important for practitioners. MVF should be prepared before an incident as part of recovery readiness, but it must be instantiated during an incident using current evidence. A preplanned MVF mission may become invalid if the required backup is contaminated, the identity system is compromised, a supplier is unavailable, or a safety interlock cannot be validated. Conversely, a different constrained mission may be viable even when full restoration is impossible.

%% file: tables/taxonomy_to_mvf_mapping_v2_4.tex
{\footnotesize
\begin{longtable}{@{}>{\RaggedRight\arraybackslash}p{0.22\textwidth} >{\RaggedRight\arraybackslash}p{0.32\textwidth} >{\RaggedRight\arraybackslash}p{0.40\textwidth}@{}}
\caption{Mapping from recovery failure modes to MVF Recovery design requirements.}\label{tab:taxonomy-mvf}\\
\toprule
Failure mode & MVF design requirement & Example control question \\
\midrule
\endfirsthead
\toprule
Failure mode & MVF design requirement & Example control question \\
\midrule
\endhead
FM01 Dependency blindness & Build mission-specific dependency graph & Which systems, data, identities, procedures, and partners are required for this product line? \\
FM02 Backup over-trust & Select restore points by evidence quality & Why is this backup clean, consistent, and suitable for the mission? \\
FM03 Identity trust collapse & Establish identity clean-room & Which privileged accounts, service accounts, and vendor access paths are trusted? \\
FM04 No proof-of-recovery & Assemble evidence bundle & What evidence supports each restoration and reconnection decision? \\
FM05 Unsafe OT reconnection & Use staged OT reintegration gates & What offline, isolated, and live checks are required before reconnection? \\
FM06 Segmentation assumption failure & Verify actual communication paths & Which paths exist now, not only on diagrams? \\
FM07 Capability mismatch & Measure recovery by production mission & What can be produced, inspected, released, and shipped? \\
FM08 Degraded mode unmanaged & Define degraded-mode limits & Which manual procedures are allowed, by whom, and until when? \\
FM09 Supplier dependency failure & Include external dependencies & Which suppliers, customers, logistics, or vendor services are required? \\
\bottomrule
\end{longtable}
}

%% file: sections/06_evidence_based_recovery_lifecycle.tex
\section{Evidence-Based Recovery Lifecycle}

\subsection{Purpose}

The lifecycle translates MVF Recovery into operational stages. It is not intended to replace existing incident response frameworks. Instead, it highlights recovery decisions that are easy to miss when responders focus on rebuilding assets. Each stage addresses one or more failure modes from Section 4.

Table~\ref{tab:lifecycle-stages} summarizes the stages and the primary failure modes addressed.

{\footnotesize
\begin{longtable}{@{}>{\RaggedRight\arraybackslash}p{0.18\textwidth} >{\RaggedRight\arraybackslash}p{0.48\textwidth} >{\RaggedRight\arraybackslash}p{0.28\textwidth}@{}}
\caption{Evidence-based recovery lifecycle stages.}\label{tab:lifecycle-stages}\\
\toprule
Stage & Primary question & Main failure modes addressed \\
\midrule
\endfirsthead
\toprule
Stage & Primary question & Main failure modes addressed \\
\midrule
\endhead
1. Mission impact assessment & What production missions are disrupted? & FM07, FM09 \\
2. Dependency modelling & What must exist together to resume a mission? & FM01, FM09 \\
3. Clean-state selection & Which restore sources and configurations can be trusted? & FM02, FM03 \\
4. MVF planning & Which constrained mission is viable now? & FM01, FM07, FM08 \\
5. Validation and simulation & Can the mission be tested before live reconnection? & FM04, FM05, FM06 \\
6. Proof-of-recovery & What evidence justifies the restart decision? & FM04 \\
7. Staged reintegration & How are systems reconnected safely? & FM05, FM06 \\
8. Monitored resumption & How is constrained production monitored and expanded? & FM07, FM08, FM09 \\
\bottomrule
\end{longtable}
}

\subsection{Stage 1: mission impact assessment}

Responders first identify which production missions are affected: product lines, batches, order types, customer commitments, quality processes, supplier flows, and logistics steps. This avoids equating system outage lists with business impact. Mission impact assessment provides the starting point for MVF selection.

\subsection{Stage 2: dependency modelling}

For each candidate mission, responders build a dependency graph covering IT systems, OT assets, identities, data, people, manual procedures, materials, suppliers, and logistics. The graph should include hidden dependencies such as service accounts, time synchronization, engineering workstations, certificate services, remote vendor access, historian feeds, and quality-release data. This stage directly addresses dependency blindness.

\subsection{Stage 3: clean-state selection}

Restore points are selected using evidence rather than recency alone. Responders assess backup age, compromise window, malware persistence, credential exposure, data consistency, dependency compatibility, and operational suitability. The decision should be documented because later production resumption depends on the trustworthiness of this choice.

\subsection{Stage 4: MVF planning}

The organization selects a constrained production mission that can be supported under current evidence. The mission may use reduced throughput, a limited product family, isolated production cell, manual scheduling, restricted remote access, or alternate supplier communication. This stage defines what “recovered enough to produce” means for the incident.

\subsection{Stage 5: validation and simulation}

Before live reconnection, responders validate the MVF plan in isolated or controlled settings where possible. Validation may include test restores, credential reset verification, configuration comparison, malware scanning, recipe validation, quality-data review, OT communication checks, and tabletop approval for degraded procedures. This stage provides evidence for trust and readiness.

\subsection{Stage 6: proof-of-recovery}

Responders assemble an evidence bundle that records what was restored, from which source, under which assumptions, with which validation results, and with which residual risk. The bundle should be understandable to technical responders, operations management, safety stakeholders, and, where relevant, auditors or regulators. It addresses FM04 by making recovery decisions reviewable.

\subsection{Stage 7: staged reintegration}

Restored systems are reintroduced through controlled gates: isolated recovery network, limited connectivity, monitoring, operator confirmation, safety approval, and incremental production use. This stage addresses FM05 and FM06 by avoiding blind reconnection based on assumed segmentation or asset availability.

\subsection{Stage 8: monitored production resumption}

MVF production begins under enhanced monitoring and explicit limits. The limits may include reduced throughput, restricted product family, manual quality checks, blocked remote access, temporary supplier procedures, or heightened logging. Exit criteria should define when to expand production, remain in degraded mode, or roll back.

\subsection{Outputs of the lifecycle}

The lifecycle should produce three concrete outputs. The first is an \textbf{MVF mission statement}, which defines the constrained production objective, duration, throughput, product scope, safety envelope, and quality-release assumptions. The second is an \textbf{MVF dependency dossier}, which lists the systems, data, identities, OT interfaces, procedures, people, suppliers, and logistics services required for the mission, together with their recovery state. The third is a \textbf{proof-of-recovery bundle}, which records evidence for clean-state selection, credential reset, configuration validation, OT reconnection approval, degraded-mode limits, and monitoring. These outputs are deliberately pragmatic. They are intended to make recovery decisions auditable and repeatable, not to impose a single universal recovery method across all factories.

\subsection{Lifecycle summary}

The lifecycle converts MVF from an analytical objective into a recovery discipline: assess missions, model dependencies, select clean states, plan constrained capability, validate recovery, document evidence, reconnect safely, and resume under monitoring. It also provides a basis for future benchmarks by defining decision points and measurable outcomes.

%% file: sections/07_benchmarking_directions.tex
\section{Benchmarking Directions and Evaluation Blueprint}

\subsection{Scope}

This paper does not release an executable benchmark. Instead, it outlines benchmarking directions for making manufacturing recovery evaluations more explicit. The goal is to avoid future work comparing recovery approaches only by asset count, backup age, or time-to-rebuild. A useful benchmark should measure whether a recovery plan restores a safe, trusted, evidence-supported production capability.

This section therefore answers RQ4 at the level of evaluation design rather than artifact release. It identifies what future datasets, simulations, tabletop exercises, and recovery-twin prototypes would need to measure.

\subsection{Scenario dimensions}

A manufacturing ransomware recovery benchmark should vary at least six dimensions. First, the production mission: product family, throughput target, quality requirements, and safety envelope. Second, the dependency graph: MES, ERP, historian, identity, engineering workstation, OT devices, supplier interface, and logistics systems. Third, compromise state: encrypted assets, credential exposure, backup contamination, lateral movement, and uncertain OT visibility. Fourth, restore options: clean, recent, incomplete, or contaminated restore points. Fifth, degraded-mode options: manual scheduling, offline quality checks, alternate supplier communication, or isolated production cell. Sixth, external constraints: supplier outage, customer deadline, regulatory requirement, or vendor access.

\subsection{Metrics}

Candidate metrics should capture capability, trust, safety, and evidence. Examples include time to MVF, percentage of required dependencies restored, number of dependency violations, number of untrusted identities reused, false-clean restore decisions, unsafe reconnection attempts, evidence completeness, degraded-mode duration, production validity, and supplier/customer coordination status. These metrics should be interpreted as scenario outcomes, not universal measures of recovery quality.

Table~\ref{tab:benchmark-metrics} lists example metric families and the recovery error each family is intended to prevent.

{\footnotesize
\begin{longtable}{@{}>{\RaggedRight\arraybackslash}p{0.22\textwidth} >{\RaggedRight\arraybackslash}p{0.36\textwidth} >{\RaggedRight\arraybackslash}p{0.36\textwidth}@{}}
\caption{Example metric families for manufacturing ransomware recovery evaluation.}\label{tab:benchmark-metrics}\\
\toprule
Metric family & Example metric & What it prevents \\
\midrule
\endfirsthead
\toprule
Metric family & Example metric & What it prevents \\
\midrule
\endhead
Capability & time to MVF; valid throughput; product-family scope & Declaring recovery based only on servers restored. \\
Dependency & dependency violations; missing critical nodes & Ignoring MES/ERP/identity/quality/supplier coupling. \\
Trust & untrusted identities reused; contaminated restore selected & Over-trusting backups or compromised credentials. \\
Safety and reintegration & unsafe reconnection attempts; failed OT gate & Reconnecting OT before validation. \\
Evidence & evidence completeness; unresolved assumptions & Restarting without proof-of-recovery. \\
Degraded operation & duration and limit violations & Allowing indefinite manual workarounds. \\
\bottomrule
\end{longtable}
}

\subsection{Baselines}

Future evaluations should compare recovery plans against simple baselines: newest-backup-first, asset-criticality-first, IT-first, OT-isolated-first, dependency-aware recovery, and evidence-aware MVF recovery. The purpose is to test whether a plan that appears faster actually restores a valid production capability or merely rebuilds convenient systems.

\subsection{Worked example blueprint}

A compact example can involve an Active Directory compromise, MES disruption, contaminated backup, historian uncertainty, and a supplier-order interface outage. A newest-backup-first plan may restore MES quickly but reuse compromised credentials and reconnect before OT validation. A dependency-aware plan may restore identity, MES, and historian in the correct order but still lack proof that the backup is clean. An evidence-aware MVF plan may take longer but produces a defensible limited-production state with clean identities, verified MES data, isolated OT reconnection, manual supplier coordination, and monitored resumption.

The point of this example is not to show that one plan is universally superior. It shows why recovery evaluation must distinguish speed from valid capability. A fast plan can be unsafe, untrusted, or operationally incomplete. A slower plan can be more valuable if it reaches a defensible production state.

Table~\ref{tab:worked-scenario} turns this blueprint into a compact evaluation example. The values are illustrative rather than experimental results; the purpose is to show how a benchmark could score competing recovery plans against the same compromised plant state.

{\footnotesize
\begin{longtable}{@{}>{\RaggedRight\arraybackslash}p{0.14\textwidth} >{\RaggedRight\arraybackslash}p{0.21\textwidth} >{\RaggedRight\arraybackslash}p{0.25\textwidth} >{\RaggedRight\arraybackslash}p{0.18\textwidth} >{\RaggedRight\arraybackslash}p{0.14\textwidth}@{}}
\caption{Worked scenario for evaluating candidate recovery plans.}\label{tab:worked-scenario}\\
\toprule
Scenario & Compromised assets & Candidate restore plan & MVF decision & Metrics outcome \\
\midrule
\endfirsthead
\toprule
Scenario & Compromised assets & Candidate restore plan & MVF decision & Metrics outcome \\
\midrule
\endhead
Single-line constrained restart for product family A. & Enterprise identity exposed; MES encrypted; newest MES backup suspected; historian integrity uncertain; supplier portal unavailable; OT cell isolated. & Newest-backup-first: restore latest MES and reconnect through existing domain credentials. & Reject MVF. Capability appears fast, but trust and evidence thresholds fail. & Low time-to-system-restore; high untrusted-identity reuse; false-clean restore risk; OT gate failure. \\
Same scenario. & Same compromised state. & Dependency-aware: rebuild identity first, restore MES, then historian, with supplier interface deferred. & Conditional MVF only if backup cleanliness and OT reconnection evidence are added. & Better dependency satisfaction; unresolved evidence completeness; delayed production validity. \\
Same scenario. & Same compromised state. & Evidence-aware MVF: reset limited operator credentials, use verified older MES restore, validate recipe offline, reconnect one OT cell through an isolated path, and use manual supplier confirmation. & Approve 48-hour MVF for one product family under monitoring and rollback limits. & Longer time-to-MVF; fewer trust violations; higher evidence completeness; valid constrained throughput. \\
\bottomrule
\end{longtable}
}

\subsection{Evaluation settings}

Future evaluations can be conducted at several levels of realism. The lowest-cost setting is a tabletop scenario in which responders reason through MVF dependencies, restore choices, and reconnection gates using a fictional but realistic plant. A second setting is a discrete-event or graph simulation that models dependency satisfaction, recovery time, uncertainty, and failure propagation. A third setting is a cyber range or OT testbed that emulates identity compromise, MES outage, historian uncertainty, and staged reconnection. The highest-realism setting is a retrospective anonymized case study using real incident timelines. Each setting trades off confidentiality, safety, cost, and validity. The blueprint is compatible with all four, but claims should be calibrated to the evaluation setting used.

\subsection{Limitations}

Benchmarking manufacturing recovery is difficult because real incident data are sensitive, safety constraints are site-specific, and production models vary. The blueprint should therefore be treated as a research agenda. Its value is to clarify assumptions and metrics, not to claim that one toy scenario can represent all factories.

%% file: sections/08_discussion_and_limitations.tex
\section{Discussion, Limitations, and Research Agenda}

\subsection{Main finding}

The main finding is that ransomware recovery in critical manufacturing should be judged by restored production capability rather than restored assets. This does not make backups, malware eradication, or system rebuilds unimportant. It means that these activities are intermediate steps toward a capability objective that includes dependencies, trust, evidence, safety, degraded operation, and external coordination.

\subsection{Why manufacturing is distinctive}

Manufacturing is distinctive because recovery decisions affect physical production and downstream supply chains. Many enterprise systems can be restored in relative isolation; factories often cannot. MES depends on identity, scheduling, recipes, quality data, historians, equipment state, operators, and supplier/customer flows. OT systems may be legacy, safety-critical, vendor-maintained, or sensitive to downtime. A recovered application can still be unusable if the quality database, engineering workstation, identity provider, supplier interface, or physical process is unavailable. This is why the introduction frames manufacturing as a specific critical-infrastructure recovery domain rather than as generic enterprise IT.

\subsection{Contribution to critical infrastructure protection}

The paper connects ransomware recovery to three critical-infrastructure themes. The first is interdependency: production capability depends on networked process components and external partners. The second is continuity: recovery should restore at least a minimum acceptable function, possibly at degraded performance. The third is safe reintegration: OT and cyber-physical assets require staged reconnection and monitoring. MVF Recovery provides an analytical objective that links these themes to actionable recovery decisions.

\subsection{Candidate patterns and under-reporting}

Weakly supported patterns were moved out of the main taxonomy but remain important research gaps. Public sources rarely describe asset-centric recovery metrics, manual playbook failure, reinfection during recovery, or quality-release blocking in enough detail to code them as core recurring failure modes. This absence is itself informative. It suggests that public incident reporting often omits the operational recovery details that researchers need to evaluate whether production was valid, safe, and sustainable.

These candidate patterns should not be ignored. Asset-centric metrics may explain why organizations declare recovery too early. Manual playbook brittleness may appear when recovery procedures do not account for degraded staffing, missing documentation, or vendor unavailability. Reinfection during recovery is widely plausible but under-documented in public manufacturing cases. Quality-release blocking is especially important because a factory may be able to produce physically but unable to release product if quality, traceability, or compliance data are unavailable. Future empirical work should treat these as targeted data-collection priorities.

\subsection{Practical implications}

For practitioners, the taxonomy can be used as a recovery readiness checklist. Before an incident, organizations can map critical production dependencies, classify restore points, define identity rebuild procedures, prepare evidence collection, document OT reconnection gates, and predefine degraded-mode limits. During an incident, MVF planning can help responders choose a constrained production mission and avoid declaring recovery based on asset restoration alone.

The most practical use of MVF is tabletop and exercise design. Instead of asking participants to restore “the MES” or “the domain,” an exercise can ask them to restore one constrained production mission under a specified compromise state. Participants must then identify dependencies, choose restore points, rebuild identity, validate OT connections, assemble evidence, and decide whether limited production is justified. This makes recovery readiness measurable without requiring disclosure of sensitive incident data.

\subsection{Research agenda}

\input{tables/research_agenda_v2_4.tex}

\subsection{Implications for recovery-twin systems}

The term Recovery Twin refers to a possible future class of decision-support systems that simulate recovery plans using dependencies, trust states, evidence, and production missions. This paper does not implement such a system. Its contribution is prior: defining the recovery objective and failure modes that a recovery twin should represent. A useful recovery twin would need to model dependency graphs, clean-state candidates, identity trust, OT reintegration gates, degraded-mode envelopes, supplier interfaces, and evidence bundles.

\subsection{Ethical, safety, and data sensitivity considerations}

Manufacturing ransomware recovery research faces data sensitivity. Detailed recovery timelines, network diagrams, supplier dependencies, and backup architectures may reveal exploitable information. Benchmarking and case reporting should therefore anonymize sensitive details while preserving recovery-relevant structure. Safety claims also require caution. This paper uses “safe” to mean operationally acceptable under available evidence and local procedures, not formally certified safety.

\subsection{Threats to validity}

The review is limited by public evidence availability, access restrictions, and reporting bias. Large incidents and public companies are more visible than smaller manufacturers. Public statements often emphasize business continuity or customer updates rather than technical recovery details. Some evidence is inference-heavy, especially for supplier and capability effects. To mitigate this, the paper uses conservative claim calibration, separates candidate gaps from main taxonomy modes, and restricts strong claims to verified full-text/source-page anchors.

There are also construct-validity risks. Terms such as “safe,” “trusted,” “clean,” and “recovered” mean different things to security teams, plant operators, auditors, and safety engineers. This paper reduces ambiguity by defining MVF success conditions, but those definitions require site-specific operationalization. External validity is also limited: pharmaceutical manufacturing, automotive assembly, food processing, semiconductor fabrication, and energy-delivery systems differ in safety, quality, and regulatory constraints. MVF is intended as a general structure, not a universal checklist.

\subsection{Boundaries}

The paper does not provide a deployed recovery system, an executable benchmark, or a formal safety verification method. It also does not claim that MVF Recovery replaces business continuity planning, incident response, or OT safety engineering. Instead, MVF complements these practices by clarifying what “recovered enough to produce” should mean after ransomware in critical manufacturing infrastructure.

%% file: tables/research_agenda_v2_4.tex
{\footnotesize
\begin{longtable}{@{}>{\RaggedRight\arraybackslash}p{0.24\textwidth} >{\RaggedRight\arraybackslash}p{0.33\textwidth} >{\RaggedRight\arraybackslash}p{0.37\textwidth}@{}}
\caption{Research agenda and future directions.}\label{tab:research-agenda}\\
\toprule
Gap & Why it matters & Future research direction \\
\midrule
\endfirsthead
\toprule
Gap & Why it matters & Future research direction \\
\midrule
\endhead
Recovery metrics remain asset-centric & Asset counts and rebuild time can overstate recovery & Develop capability-based recovery metrics aligned with MVF missions. \\
Full incident timelines are rarely public & Recovery sequencing cannot be validated from high-level statements & Build anonymized manufacturing recovery case repositories. \\
Quality-release blocking is under-reported & Production may be physically possible but commercially invalid & Study quality, traceability, laboratory, and regulatory dependencies after ransomware. \\
OT reintegration evidence is limited & Unsafe reconnection can create safety and process risks & Evaluate staged OT reconnection gates in testbeds and tabletop exercises. \\
Identity recovery is under-modelled & Compromised identity can invalidate otherwise restored systems & Develop clean-room identity recovery protocols for IT/OT environments. \\
Supplier recovery coordination is weakly formalized & Manufacturing disruption cascades across firms & Model supplier/customer coordination as part of MVF dependency graphs. \\
Benchmarks lack recovery semantics & Recovery tools are hard to compare & Develop benchmark scenarios that measure time to trusted capability, not only time to restore. \\
Recovery-twin systems are conceptual & Decision support needs explicit assumptions & Prototype recovery twins using dependency, trust, evidence, and mission models. \\
\bottomrule
\end{longtable}
}

%% file: sections/09_conclusion.tex
\section{Conclusion}

Ransomware recovery in critical manufacturing infrastructure is a capability-restoration problem. A factory can rebuild servers, restore backups, or reconnect applications while still being unable to produce safely, validate quality, authenticate operators, coordinate suppliers, or prove that recovered systems are trustworthy. This paper synthesized academic, guidance, incident, and verified full-text/source-page evidence to identify nine evidence-backed recovery failure modes: dependency blindness, backup over-trust, identity trust collapse, no proof-of-recovery, unsafe OT reconnection, segmentation assumption failure, capability mismatch, unmanaged degraded operation, and supplier dependency failure.

The paper then introduced Minimum Viable Factory Recovery as an analytical objective for resuming the smallest safe, trusted, and operationally meaningful production capability under current dependency, evidence, identity, data, network, OT, and supplier constraints. MVF Recovery shifts the recovery question from “which assets are back?” to “which constrained production mission can be validly resumed?” The derived lifecycle and benchmarking directions provide a foundation for future recovery evaluation, recovery-twin research, and practitioner readiness work. The central message is simple: in manufacturing, ransomware recovery should be proved through production capability, dependency integrity, trust reconstruction, and safe reintegration, not through asset restoration alone.

%% file: sections/10_references.tex
[1] A. Kharraz, W. Robertson, D. Balzarotti, L. Bilge, and E. Kirda, "Cutting the Gordian knot: A look under the hood of ransomware attacks," in \textit{Proc. 12th International Conference on Detection of Intrusions and Malware, and Vulnerability Assessment (DIMVA)}, 2015, pp. 3--24, doi: 10.1007/978-3-319-20550-2\_1.

[2] H. Oz, A. Aris, A. Levi, and A. S. Uluagac, "A survey on ransomware: Evolution, taxonomy, and defense solutions," \textit{ACM Computing Surveys}, vol. 54, no. 11s, pp. 1--37, 2022, doi: 10.1145/3514229.

[3] C. Beaman, A. Barkworth, T. D. Akande, S. Hakak, and M. K. Khan, "Ransomware: Recent advances, analysis, challenges and future research directions," \textit{Computers \& Security}, vol. 111, Art. no. 102490, 2021, doi: 10.1016/j.cose.2021.102490.

[4] Cybersecurity and Infrastructure Security Agency, Federal Bureau of Investigation, National Security Agency, and Multi-State Information Sharing and Analysis Center, \textit{\#StopRansomware Guide}, updated Oct. 19, 2023. [Online]. Available: \url{https://www.cisa.gov/stopransomware/ransomware-guide}. Accessed: Apr. 29, 2026.

[5] B. Fisher, M. Souppaya, W. Barker, and K. Scarfone, \textit{Ransomware Risk Management: A Cybersecurity Framework Profile}, NISTIR 8374, National Institute of Standards and Technology, Feb. 2022, doi: 10.6028/NIST.IR.8374.

[6] National Cyber Security Centre, "Mitigating malware and ransomware attacks," guidance, updated guidance page. [Online]. Available: \url{https://www.ncsc.gov.uk/guidance/mitigating-malware-and-ransomware-attacks}. Accessed: Apr. 29, 2026.

[7] M. Benmalek, "Ransomware on cyber-physical systems: Taxonomies, case studies, security gaps, and open challenges," \textit{Internet of Things and Cyber-Physical Systems}, vol. 4, pp. 186--202, 2024, doi: 10.1016/j.iotcps.2023.12.001.

[8] K. Stouffer, M. Pease, C. Y. Tang, T. Zimmerman, V. Pillitteri, S. Lightman, A. Hahn, S. Saravia, A. Sherule, and M. Thompson, \textit{Guide to Operational Technology (OT) Security}, NIST Special Publication 800-82 Rev. 3, National Institute of Standards and Technology, Sep. 2023, doi: 10.6028/NIST.SP.800-82r3.

[9] MITRE, "ATT\&CK for ICS matrix," MITRE ATT\&CK knowledge base. [Online]. Available: \url{https://attack.mitre.org/matrices/ics/}. Accessed: Apr. 29, 2026.

[10] MITRE, "Inhibit response function, Tactic TA0107," MITRE ATT\&CK for ICS. [Online]. Available: \url{https://attack.mitre.org/tactics/TA0107/}. Accessed: Apr. 29, 2026.

[11] MITRE, "Impair process control, Tactic TA0106," MITRE ATT\&CK for ICS. [Online]. Available: \url{https://attack.mitre.org/tactics/TA0106/}. Accessed: Apr. 29, 2026.

[12] MITRE, "Impact, Tactic TA0105," MITRE ATT\&CK for ICS. [Online]. Available: \url{https://attack.mitre.org/tactics/TA0105/}. Accessed: Apr. 29, 2026.

[13] Y. Zhang, Z. Sun, L. Yang, Z. Li, Q. Zeng, Y. He, and X. Zhang, "All your PLCs belong to me: ICS ransomware is realistic," in \textit{Proc. IEEE 19th International Conference on Trust, Security and Privacy in Computing and Communications (TrustCom)}, 2020, pp. 502--509, doi: 10.1109/TrustCom50675.2020.00074.

[14] M. Al-Hawawreh, F. den Hartog, and E. Sitnikova, "Targeted ransomware: A new cyber threat to edge system of brownfield Industrial Internet of Things," \textit{IEEE Internet of Things Journal}, vol. 6, no. 4, pp. 7137--7151, 2019, doi: 10.1109/JIOT.2019.2914390.

[15] D. M. Nicol, "The ransomware threat to energy-delivery systems," \textit{IEEE Security \& Privacy}, vol. 19, no. 3, pp. 24--32, 2021, doi: 10.1109/MSEC.2021.3063678.

[16] R. Pal, R. X. Sequeira, S. Zeijlmaker, and M. Siegel, "Optimizing cyber-resilience in critical infrastructure networks," in \textit{Proc. 2024 Winter Simulation Conference (WSC)}, 2024, pp. 774--785, doi: 10.1109/WSC63780.2024.10838999.

[17] A. Singh, A. R. Ikuesan, and H. S. Venter, "Digital forensic readiness framework for ransomware investigation," in \textit{Digital Forensics and Cyber Crime: 10th International EAI Conference, ICDF2C 2018}, Lecture Notes of the Institute for Computer Sciences, Social Informatics and Telecommunications Engineering, Springer, 2019, pp. 91--105, doi: 10.1007/978-3-030-05487-8\_5.

[18] P. Nakhonthai and K. Chimmanee, "Digital forensic analysis of ransomware attacks on industrial control systems: A case study in factories," in \textit{Proc. 2022 6th International Conference on Information Technology (InCIT)}, 2022, pp. 416--421, doi: 10.1109/InCIT56086.2022.10067356.

[19] P. Bajpai, \textit{Extracting Ransomware's Keys by Utilizing Memory Forensics}, Ph.D. dissertation, Michigan State University, 2020, ProQuest no. 27837280.

[20] A. Cartwright and E. Cartwright, "The economics of ransomware attacks on integrated supply chain networks," \textit{Digital Threats: Research and Practice}, vol. 4, no. 4, 2023, doi: 10.1145/3579647.

[21] R. Pergande, J. Hamann-Lohmer, and R. Lasch, "From attack to adaptation: A case study of capabilities driving digital supply chain recovery," \textit{IEEE Engineering Management Review}, early access, 2025, doi: 10.1109/EMR.2025.3568586.

[22] L. Manning and A. Kowalska, "The threat of ransomware in the food supply chain: A challenge for food defence," \textit{Trends in Organized Crime}, 2023, doi: 10.1007/s12117-023-09516-y.

[23] T. Miller, A. Staves, S. Maesschalck, M. Sturdee, and B. Green, "Looking back to look forward: Lessons learnt from cyber-attacks on industrial control systems," \textit{International Journal of Critical Infrastructure Protection}, vol. 35, Art. no. 100464, 2021, doi: 10.1016/j.ijcip.2021.100464.

[24] A. Staves, T. Anderson, A. Balderstone, B. Green, A. Gouglidis, and D. Hutchison, "A cyber incident response and recovery framework to support operators of industrial control systems," \textit{International Journal of Critical Infrastructure Protection}, vol. 37, Art. no. 100505, 2022, doi: 10.1016/j.ijcip.2022.100505.

[25] J. Huang, J. Xu, X. Xing, P. Liu, and M. K. Qureshi, "FlashGuard: Leveraging intrinsic flash properties to defend against encryption ransomware," in \textit{Proc. ACM SIGSAC Conference on Computer and Communications Security (CCS)}, 2017, pp. 2231--2244, doi: 10.1145/3133956.3134035.

[26] J. Dafoe, N. Chen, B. Chen, and Z. Wang, "Enabling per-file data recovery from ransomware attacks via file system forensics and flash translation layer data extraction," \textit{Cybersecurity}, vol. 7, Art. no. 75, 2024, doi: 10.1186/s42400-024-00287-9.

[27] A. Nelson, S. Rekhi, M. Souppaya, and K. Scarfone, \textit{Incident Response Recommendations and Considerations for Cybersecurity Risk Management: A CSF 2.0 Community Profile}, NIST Special Publication 800-61 Rev. 3, National Institute of Standards and Technology, Apr. 2025, doi: 10.6028/NIST.SP.800-61r3.

[28] H. Hmiddouch, A. Villafranca, R. Castro, V. Dubetskyy, and M.-D. Cano, "Enhancing industrial cybersecurity with virtual lab simulations," \textit{International Journal of Advanced Computer Science and Applications}, vol. 16, no. 5, pp. 40--50, 2025.

[29] U. J. Butt, M. Abbod, A. Lors, H. Jahankhani, A. Jamal, and A. Kumar, "Ransomware threat and its impact on SCADA," in \textit{Proc. 12th International Conference on Global Security, Safety and Sustainability (ICGS3)}, 2019, doi: 10.1109/ICGS3.2019.8688327.

[30] J. Ibarra, U. J. Butt, A. Do, H. Jahankhani, and A. Jamal, "Ransomware impact to SCADA systems and its scope to critical infrastructure," in \textit{Proc. 2019 IEEE 12th International Conference on Global Security, Safety and Sustainability (ICGS3)}, 2019, pp. 1--12, doi: 10.1109/ICGS3.2019.8688299.

[31] N. Suk-on, N. Thiratitsakun, and K. Chimmanee, "Digital forensic analysis of LockBit ransomware attack on operational technology," in \textit{Proc. 8th International Conference on Information Technology (InCIT)}, 2024, pp. 624--629, doi: 10.1109/InCIT63192.2024.10810564.

[32] Y. Ota, E. Mizuno, K. Watarai, T. Aoyama, T. Hamaguchi, Y. Hashimoto, and I. Koshijima, "Development of a hybrid exercise for organizational cyber resilience," in \textit{Safety and Security Engineering IX}, WIT Transactions on the Built Environment, vol. 206, WIT Press, 2021, pp. 55--65, doi: 10.2495/SAFE210051.

[33] M. Athinaiou, H. Mouratidis, T. Fotis, M. Pavlidis, and E. Panaousis, "Towards the definition of a security incident response modelling language," in \textit{Trust, Privacy and Security in Digital Business}, LNCS 11033, Springer, 2018, pp. 198--212, doi: 10.1007/978-3-319-98385-1\_14.

[34] A. J. Staves, \textit{Operational Technology Preparedness: A Risk-Based Safety Approach to Scoping Security Tests for Cyber Incident Response and Recovery}, Ph.D. dissertation, Lancaster University, 2023, doi: 10.17635/lancaster/thesis/2111.

[35] T. N. I. Alrumaih, M. J. F. Alenazi, N. A. AlSowaygh, A. A. Humayed, and I. A. Alablani, "Cyber resilience in industrial networks: A state of the art, challenges, and future directions," \textit{Journal of King Saud University - Computer and Information Sciences}, vol. 35, no. 9, Art. no. 101781, 2023, doi: 10.1016/j.jksuci.2023.101781.

[36] H. Harkat, L. M. Camarinha-Matos, J. Goes, and H. F. T. Ahmed, "Cyber-physical systems security: A systematic review," \textit{Computers \& Industrial Engineering}, vol. 188, Art. no. 109891, 2024, doi: 10.1016/j.cie.2024.109891.

[37] M. Rahman and M. S. Shafae, "Cyber-physical security vulnerabilities identification and classification in smart manufacturing," arXiv preprint, 2025.

[38] M. Akbarzadeh and S. Katsikas, "Dependency-based security risk assessment for cyber-physical systems," \textit{International Journal of Information Security}, 2023.

[39] S. M. Khalil, H. Bahsi, and T. Korõtko, "Threat modeling of industrial control systems: A systematic literature review," \textit{Computers \& Security}, vol. 136, Art. no. 103543, 2024, doi: 10.1016/j.cose.2023.103543.

[40] M.-C. Ilau, A. Baldwin, T. Caulfield, and D. Pym, "Modelling and simulating organizational ransomware recovery: Structure, methodology, and decisions," \textit{Journal of Cybersecurity}, vol. 11, no. 1, Art. no. tyaf035, 2025, doi: 10.1093/cybsec/tyaf035.

[41] JBS USA, "JBS USA and Pilgrim's announce resolution of cyberattack," company press release, Jun. 3, 2021. [Online]. Available: \url{https://jbsfoodsgroup.com/articles/jbs-usa-and-pilgrim-s-announce-resolution-of-cyberattack}. Accessed: Apr. 29, 2026.

[42] JBS USA, "JBS USA cyberattack media statement - June 9," company press release, Jun. 9, 2021. [Online]. Available: \url{https://jbsfoodsgroup.com/articles/jbs-usa-cyberattack-media-statement-june-9}. Accessed: Apr. 29, 2026.

[43] Jaguar Land Rover Automotive plc, "Statement on cyber incident," JLR Media Newsroom, Sep. 29, 2025. [Online]. Available: \url{https://media.jaguarlandrover.com/news/2025/09/statement-cyber-incident-6}. Accessed: Apr. 29, 2026.

[44] Asahi Group Holdings, Ltd., "Update on system disruption due to cyberattack (2nd)," Newsroom, Oct. 3, 2025. [Online]. Available: \url{https://www.asahigroup-holdings.com/en/newsroom/detail/20251003-0204.html}. Accessed: Apr. 29, 2026.

[45] J. L., "Offline backups in an online world," National Cyber Security Centre blog, 2017. [Online]. Available: \url{https://www.ncsc.gov.uk/blog-post/offline-backups-in-an-online-world}. Accessed: Apr. 29, 2026.

[46] European Union Agency for Cybersecurity, \textit{ENISA Threat Landscape 2024}, ENISA, 2024. [Online]. Available: \url{https://www.enisa.europa.eu/publications/enisa-threat-landscape-2024}. Accessed: Apr. 29, 2026.

[47] Kaspersky ICS CERT, "A brief overview of the main incidents in industrial cybersecurity: Q1 2025," Kaspersky ICS CERT, Jun. 26, 2025. [Online]. Available: \url{https://ics-cert.kaspersky.com/publications/reports/2025/06/26/a-brief-overview-of-the-main-incidents-in-industrial-cybersecurity-q1-2025/}. Accessed: Apr. 29, 2026.

[48] Dragos, "Dragos's 8th annual OT cybersecurity year in review is now available," Dragos Blog, 2025. [Online]. Available: \url{https://www.dragos.com/blog/dragos-8th-annual-ot-cybersecurity-year-in-review-is-now-available}. Accessed: Apr. 29, 2026.

[49] Google Cloud/Mandiant, \textit{M-Trends 2025}, Google Cloud, 2025. [Online]. Available: \url{https://cloud.google.com/security/resources/m-trends}. Accessed: Apr. 29, 2026.

[50] Microsoft, "Microsoft defense against ransomware, extortion, and intrusion," Microsoft Learn. [Online]. Available: \url{https://learn.microsoft.com/en-us/security/ransomware/}. Accessed: Apr. 29, 2026.

[51] Cybersecurity and Infrastructure Security Agency, \textit{Cross-Sector Cybersecurity Performance Goals}, CISA. [Online]. Available: \url{https://www.cisa.gov/cross-sector-cybersecurity-performance-goals}. Accessed: Apr. 29, 2026.

[52] C. Pascoe, S. Quinn, and K. Scarfone, \textit{The NIST Cybersecurity Framework (CSF) 2.0}, NIST Cybersecurity White Paper 29, National Institute of Standards and Technology, Feb. 2024, doi: 10.6028/NIST.CSWP.29.

[53] International Electrotechnical Commission, \textit{IEC 62443: Industrial communication networks -- Network and system security}, IEC 62443 series, Geneva, Switzerland.

[54] D. Dolev and A. C. Yao, "On the security of public key protocols," \textit{IEEE Transactions on Information Theory}, vol. 29, no. 2, pp. 198--208, 1983, doi: 10.1109/TIT.1983.1056650.

[55] B. Kitchenham and S. Charters, \textit{Guidelines for Performing Systematic Literature Reviews in Software Engineering}, EBSE Technical Report EBSE-2007-01, Keele University and Durham University, 2007.

[56] M. J. Page et al., "The PRISMA 2020 statement: An updated guideline for reporting systematic reviews," \textit{BMJ}, vol. 372, Art. no. n71, 2021, doi: 10.1136/bmj.n71.

[57] V. Garousi, M. Felderer, and M. V. Mantyla, "Guidelines for including grey literature and conducting multivocal literature reviews in software engineering," \textit{Information and Software Technology}, vol. 106, pp. 101--121, 2019, doi: 10.1016/j.infsof.2018.09.006.

[58] K. Petersen, R. Feldt, S. Mujtaba, and M. Mattsson, "Systematic mapping studies in software engineering," in \textit{Proc. 12th International Conference on Evaluation and Assessment in Software Engineering (EASE)}, 2008, pp. 68--77.

[59] C. Wohlin, "Guidelines for snowballing in systematic literature studies and a replication in software engineering," in \textit{Proc. 18th International Conference on Evaluation and Assessment in Software Engineering (EASE)}, 2014, Art. no. 38, doi: 10.1145/2601248.2601268.

[60] Joint Task Force, \textit{Security and Privacy Controls for Information Systems and Organizations}, NIST Special Publication 800-53 Rev. 5, National Institute of Standards and Technology, Sep. 2020, doi: 10.6028/NIST.SP.800-53r5.

[61] S. Rose, O. Borchert, S. Mitchell, and S. Connelly, \textit{Zero Trust Architecture}, NIST Special Publication 800-207, National Institute of Standards and Technology, Aug. 2020, doi: 10.6028/NIST.SP.800-207.

[62] M. Swanson, P. Bowen, A. W. Phillips, D. Gallup, and D. Lynes, \textit{Contingency Planning Guide for Federal Information Systems}, NIST Special Publication 800-34 Rev. 1, National Institute of Standards and Technology, May 2010, doi: 10.6028/NIST.SP.800-34r1.

[63] P. A. Grassi, M. E. Garcia, and J. L. Fenton, \textit{Digital Identity Guidelines}, NIST Special Publication 800-63-3, National Institute of Standards and Technology, Jun. 2017, doi: 10.6028/NIST.SP.800-63-3.

[64] Cybersecurity and Infrastructure Security Agency, "Known Exploited Vulnerabilities Catalog," CISA. [Online]. Available: \url{https://www.cisa.gov/known-exploited-vulnerabilities-catalog}. Accessed: Apr. 29, 2026.

[65] National Cyber Security Centre, "Small Business Guide: Response and recovery," NCSC guidance. [Online]. Available: \url{https://www.ncsc.gov.uk/collection/small-business-guide/response-and-recovery}. Accessed: Apr. 29, 2026.

[66] A. Aljoghaiman and V. P. K. Sundram, "Mitigating ransomware risks in manufacturing and the supply chain: A comprehensive security framework," \textit{International Journal of Cyber Criminology}, vol. 17, no. 2, pp. 231--249, 2023, doi: 10.5281/zenodo.4766714.

[67] S. Maesschalck, A. Staves, R. Derbyshire, B. Green, and D. Hutchison, "Walking under the ladder logic: PLC-VBS: A PLC control logic vulnerability scanning tool," \textit{Computers \& Security}, vol. 127, Art. no. 103116, 2023, doi: 10.1016/j.cose.2023.103116.

[68] M. Musluoglu, N. Kunicina, and J. Caiko, "Vulnerability assessment of industrial control systems for Colonial Pipeline and WannaCry ransomware," in \textit{Proc. IEEE 65th International Scientific Conference on Power and Electrical Engineering of Riga Technical University (RTUCON)}, 2024, doi: 10.1109/RTUCON62997.2024.10830848.

[69] Norsk Hydro ASA, "Cyber attack on Hydro," 2019. [Online]. Available: \url{https://www.hydro.com/en/global/media/on-the-agenda/cyber-attack/}. Accessed: Apr. 29, 2026.

[70] Norsk Hydro ASA, \textit{Annual Report 2019}, 2020. [Online]. Available: \url{https://www.hydro.com/Document/Doc/Annual\%20report\%202019\%20web.pdf?docId=506433}. Accessed: Apr. 29, 2026.

[71] Toyota Motor Corporation, "Deepening ties in difficult times: One year on from the Kojima Industries cyberattack," \textit{Toyota Times}, 2023. [Online]. Available: \url{https://toyotatimes.jp/en/newscast/008.html}. Accessed: Apr. 29, 2026.

[72] MKS Instruments, Inc., "MKS Instruments provides update on ransomware event," Form 8-K, Exhibit 99.1, U.S. Securities and Exchange Commission, Feb. 2023. [Online]. Available: \url{https://www.sec.gov/Archives/edgar/data/1049502/000119312523034334/d464518dex991.htm}. Accessed: Apr. 29, 2026.

[73] A.P. Moller - Maersk A/S, "Cyber attack update," investor release, Jun. 2017. [Online]. Available: \url{https://investor.maersk.com/node/19831/pdf}. Accessed: Apr. 29, 2026.

[74] The Clorox Company, "Form 8-K: Cybersecurity incident disclosure," U.S. Securities and Exchange Commission, Aug. 14, 2023. [Online]. Available: \url{https://www.sec.gov/Archives/edgar/data/21076/000120677423001133/clx4242401-8k.htm}. Accessed: Apr. 29, 2026.

[75] Dole plc, "Dole experiences cybersecurity incident," company press release, Feb. 22, 2023. [Online]. Available: \url{https://www.dole.com/press/2023/dole-experiences-cybersecurity-incident}. Accessed: Apr. 29, 2026.

[76] Z. Whittaker, "Honda's global operations halted by ransomware attack," \textit{TechCrunch}, Jun. 9, 2020. [Online]. Available: \url{https://techcrunch.com/2020/06/09/honda-ransomware-snake/}. Accessed: Apr. 29, 2026.